\documentclass[nofootinbib,twocolumn,showpacs,prl,superscriptaddress,secnumarabic,amssymb,nobibnotes,aps,floatfix]{revtex4-1}
\usepackage{graphicx}
\usepackage{hyperref}
\begin{document}

\title{Cross sections for the exclusive photon electroproduction on
the proton and Generalized Parton Distributions}

\newcommand*{\ORSAY}{Institut de Physique Nucl\'eaire, CNRS/IN2P3 and Universit\'e Paris Sud, Orsay, France}
\newcommand*{\ORSAYindex}{1}
\affiliation{\ORSAY}
\newcommand*{\JLAB}{Thomas Jefferson National Accelerator Facility, Newport News, Virginia 23606}
\newcommand*{\JLABindex}{2}
\affiliation{\JLAB}
\newcommand*{\SACLAY}{CEA, Centre de Saclay, Irfu/Service de Physique Nucl\'eaire, 91191 Gif-sur-Yvette, France}
\newcommand*{\SACLAYindex}{3}
\affiliation{\SACLAY}
\newcommand*{\RPI}{Rensselaer Polytechnic Institute, Troy, New York 12180-3590}
\newcommand*{\RPIindex}{4}
\affiliation{\RPI}
\newcommand*{\ANL}{Argonne National Laboratory, Argonne, Illinois 60439}
\newcommand*{\ANLindex}{5}
\affiliation{\ANL}
\newcommand*{\ASU}{Arizona State University, Tempe, Arizona 85287-1504}
\newcommand*{\ASUindex}{6}
\affiliation{\ASU}
\newcommand*{\CSUDH}{California State University, Dominguez Hills, Carson, CA 90747}
\newcommand*{\CSUDHindex}{7}
\affiliation{\CSUDH}
\newcommand*{\CANISIUS}{Canisius College, Buffalo, NY}
\newcommand*{\CANISIUSindex}{8}
\affiliation{\CANISIUS}
\newcommand*{\CMU}{Carnegie Mellon University, Pittsburgh, Pennsylvania 15213}
\newcommand*{\CMUindex}{9}
\affiliation{\CMU}
\newcommand*{\CUA}{Catholic University of America, Washington, D.C. 20064}
\newcommand*{\CUAindex}{10}
\affiliation{\CUA}
\newcommand*{\UCONN}{University of Connecticut, Storrs, Connecticut 06269}
\newcommand*{\UCONNindex}{11}
\affiliation{\UCONN}
\newcommand*{\FU}{Fairfield University, Fairfield CT 06824}
\newcommand*{\FUindex}{12}
\affiliation{\FU}
\newcommand*{\FIU}{Florida International University, Miami, Florida 33199}
\newcommand*{\FIUindex}{13}
\affiliation{\FIU}
\newcommand*{\FSU}{Florida State University, Tallahassee, Florida 32306}
\newcommand*{\FSUindex}{14}
\affiliation{\FSU}
\newcommand*{\GWUI}{The George Washington University, Washington, DC 20052}
\newcommand*{\GWUIindex}{15}
\affiliation{\GWUI}
\newcommand*{\ISU}{Idaho State University, Pocatello, Idaho 83209}
\newcommand*{\ISUindex}{16}
\affiliation{\ISU}
\newcommand*{\INFNFE}{INFN, Sezione di Ferrara, 44100 Ferrara, Italy}
\newcommand*{\INFNFEindex}{17}
\affiliation{\INFNFE}
\newcommand*{\INFNFR}{INFN, Laboratori Nazionali di Frascati, 00044 Frascati, Italy}
\newcommand*{\INFNFRindex}{18}
\affiliation{\INFNFR}
\newcommand*{\INFNGE}{INFN, Sezione di Genova, 16146 Genova, Italy}
\newcommand*{\INFNGEindex}{19}
\affiliation{\INFNGE}
\newcommand*{\INFNRO}{INFN, Sezione di Roma Tor Vergata, 00133 Rome, Italy}
\newcommand*{\INFNROindex}{20}
\affiliation{\INFNRO}
\newcommand*{\INFNTUR}{INFN, Sezione di Torino, 10125 Torino, Italy}
\newcommand*{\INFNTURindex}{21}
\affiliation{\INFNTUR}
\newcommand*{\ITEP}{Institute of Theoretical and Experimental Physics, Moscow, 117259, Russia}
\newcommand*{\ITEPindex}{22}
\affiliation{\ITEP}
\newcommand*{\JMU}{James Madison University, Harrisonburg, Virginia 22807}
\newcommand*{\JMUindex}{23}
\affiliation{\JMU}
\newcommand*{\KNU}{Kyungpook National University, Daegu 702-701, Republic of Korea}
\newcommand*{\KNUindex}{24}
\affiliation{\KNU}
\newcommand*{\UNH}{University of New Hampshire, Durham, New Hampshire 03824-3568}
\newcommand*{\UNHindex}{25}
\affiliation{\UNH}
\newcommand*{\NSU}{Norfolk State University, Norfolk, Virginia 23504}
\newcommand*{\NSUindex}{26}
\affiliation{\NSU}
\newcommand*{\OHIOU}{Ohio University, Athens, Ohio  45701}
\newcommand*{\OHIOUindex}{27}
\affiliation{\OHIOU}
\newcommand*{\ODU}{Old Dominion University, Norfolk, Virginia 23529}
\newcommand*{\ODUindex}{28}
\affiliation{\ODU}
\newcommand*{\ROMAII}{Universit\`{a} di Roma Tor Vergata, 00133 Rome Italy}
\newcommand*{\ROMAIIindex}{29}
\affiliation{\ROMAII}
\newcommand*{\MSU}{Skobeltsyn Institute of Nuclear Physics, Lomonosov Moscow State University, 119234 Moscow, Russia}
\newcommand*{\MSUindex}{30}
\affiliation{\MSU}
\newcommand*{\SCAROLINA}{University of South Carolina, Columbia, South Carolina 29208}
\newcommand*{\SCAROLINAindex}{31}
\affiliation{\SCAROLINA}
\newcommand*{\TEMPLE}{Temple University,  Philadelphia, PA 19122 }
\newcommand*{\TEMPLEindex}{32}
\affiliation{\TEMPLE}
\newcommand*{\UTFSM}{Universidad T\'{e}cnica Federico Santa Mar\'{i}a, Casilla 110-V Valpara\'{i}so, Chile}
\newcommand*{\UTFSMindex}{33}
\affiliation{\UTFSM}
\newcommand*{\EDINBURGH}{Edinburgh University, Edinburgh EH9 3JZ, United Kingdom}
\newcommand*{\EDINBURGHindex}{34}
\affiliation{\EDINBURGH}
\newcommand*{\GLASGOW}{University of Glasgow, Glasgow G12 8QQ, United Kingdom}
\newcommand*{\GLASGOWindex}{35}
\affiliation{\GLASGOW}
\newcommand*{\VT}{Virginia Tech, Blacksburg, Virginia   24061-0435}
\newcommand*{\VTindex}{36}
\affiliation{\VT}
\newcommand*{\VIRGINIA}{University of Virginia, Charlottesville, Virginia 22901}
\newcommand*{\VIRGINIAindex}{37}
\affiliation{\VIRGINIA}
\newcommand*{\WM}{College of William and Mary, Williamsburg, Virginia 23187-8795}
\newcommand*{\WMindex}{38}
\affiliation{\WM}
\newcommand*{\YEREVAN}{Yerevan Physics Institute, 375036 Yerevan, Armenia}
\newcommand*{\YEREVANindex}{39}
\affiliation{\YEREVAN}
 
\newcommand*{\NOWJLAB}{Thomas Jefferson National Accelerator Facility, Newport News, Virginia 23606}
\newcommand*{\NOWODU}{Old Dominion University, Norfolk, Virginia 23529}

\author {H.S.~Jo}
\email[corresponding author: ]{jo@ipno.in2p3.fr}
\affiliation{\ORSAY}
\author {F.X.~Girod} 
\affiliation{\JLAB}
\affiliation{\SACLAY}
\author {H.~Avakian} 
\affiliation{\JLAB}
\author {V.D.~Burkert}
\affiliation{\JLAB}
\author {M.~Gar\c{c}on}
\affiliation{\SACLAY}
\author {M.~Guidal}
\affiliation{\ORSAY}
\author {V.~Kubarovsky} 
\affiliation{\JLAB}
\affiliation{\RPI}
\author {S.~Niccolai}
\affiliation{\ORSAY}
\author {P.~Stoler} 
\affiliation{\RPI}
\author {K.P.~Adhikari} 
\affiliation{\ODU}
% ...

\collaboration{The CLAS Collaboration}
\noaffiliation

\date{\today}
\begin{abstract}
Unpolarized and beam-polarized four-fold cross sections
$\frac{d^4 \sigma}{dQ^2 dx_B dt d\phi}$
for the $ep\to e^\prime p^\prime \gamma$ reaction were measured
using the CLAS detector and the 5.75-GeV polarized electron
beam of the Jefferson Lab accelerator, for 110 ($Q^2,x_B,t$) bins over the widest
phase space ever explored in the valence-quark region.
Several models of Generalized Parton Distributions (GPDs) describe the data well
at most of our kinematics.
This increases our confidence that we understand the GPD $H$, expected to be the dominant contributor to these
observables.
Through a leading-twist extraction of Compton Form Factors, these results reveal
a tomographic image of the nucleon.
\end{abstract}
\pacs{12.38.-t, 13.40.Gp, 13.60.Fz, 13.60.Hb, 14.20.Dh, 24.85.+p}

\maketitle 

The internal structure and dynamics of the proton, the nucleus of the most abundant
chemical element in the visible universe, 
still remain a mystery in many respects,
more than 40 years after the evidence for its quark and gluon substructure.
How are the spatial and momentum distributions of the quarks and gluons (\textit{i.e.},
the partons) correlated inside the nucleon? How do the partons contribute
to the bulk properties of the proton (mass, spin, charge,...)? These are some fundamental
questions at the intersection of nuclear and particle physics which are still to be resolved.

In order to tackle these essential issues, a large experimental program was launched
worldwide at Jefferson Lab (JLab), COMPASS and HERA, facilities using multi-GeV
electromagnetic probes, to study deeply virtual Compton
scattering (DVCS). In the valence-quark region, this corresponds to Compton scattering
at the quark level, with the incoming photon radiated from the lepton beam. As in the study of
atomic or nuclear structure, the energy and angular distributions of the scattered photon reflect
the distribution in momentum and/or space of the targets, which in our case are the quarks
inside the proton. At JLab, electron beams are used and the reaction to study proton
structure is $ep\to e^\prime p^\prime \gamma$.
It was shown~\cite{Mueller:1998fv,Ji:1996ek,Ji:1996nm,Radyushkin:1996nd,Radyushkin:1997ki}
that this process, at sufficiently large squared electron momentum transfer 
$Q^2=-(e-e^\prime)^2$ 
and small squared proton momentum transfer $t=(p-p^\prime)^2$ (in terms of the electron
and proton four-vectors), could be interpreted in the framework
of Quantum Chromodynamics (QCD), the fundamental theory governing the interaction of quarks and 
gluons, as the product of the elementary Compton scattering at the quark level $\gamma^* q\to 
\gamma q$ with factorizable structure functions called Generalized Parton Distributions (GPDs).
\begin{figure}[tb]
\includegraphics[width=8.9cm]{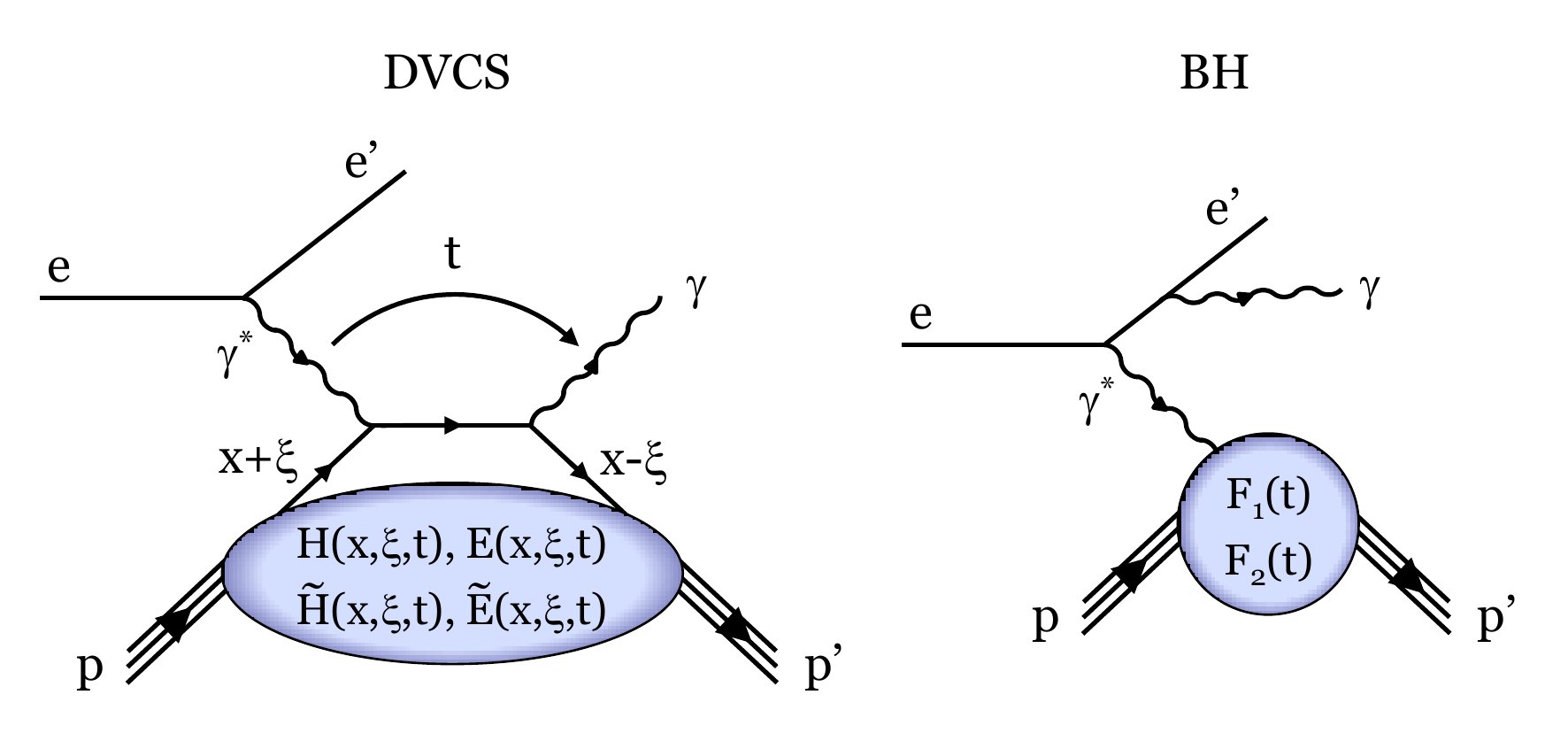}
\vspace{-0.9cm}
\caption{Left: the dominant mechanism for the DVCS process at sufficiently large $Q^2$ and 
small $|t|$, as predicted by the QCD factorization theorem.
Right: the Bethe-Heitler (BH) process, which interferes with the DVCS process in the
$ep\to e^\prime p^\prime \gamma$ reaction.}
\label{fig:diags}
\end{figure}

Figure~\ref{fig:diags} (left) illustrates GPD QCD factorization for the DVCS process.
In a frame where the nucleon moves at the speed of light in a given direction, a quark with 
longitudinal momentum fraction $x+\xi$ absorbs the virtual photon and, after radiating the
final-state photon, the \textit{same} quark returns into the nucleon with a longitudinal momentum
fraction $x-\xi$, plus some transverse kick included in $t$. The GPDs are functions of
$x$, $\xi$, and $t$, and represent the probability amplitude of such a process.
The variable $\xi$ is related to the
Bjorken variable $x_B$: $\xi\approx {{x_B}\over{2-x_B}}$, where $x_B=\frac{Q^2}{2M\nu}$ with
the proton mass $M$ and $\nu=E_e-E_{e^\prime}$.
Thus, it is determined by the scattered-electron kinematics. The quantity $x$ is
not measurable in the DVCS process.
At leading-order QCD, GPDs do not depend on $Q^2$. At leading-twist QCD, four GPDs enter the
description of the DVCS process: $H$, $\tilde H$, $E$ and $\tilde E$, representing the four
independent helicity-spin transitions of the quark-nucleon system between the initial and
final states. The GPDs are QCD matrix elements that integrate and project on a few variables
the full complexity of the quarks' and gluons' dynamics within the nucleon.

The GPDs embody the longitudinal momentum distribution of the quarks in the nucleon,
their transverse spatial distribution, and the correlation between these two distributions.
One uses the term \textit{nucleon tomography} as
one can probe the transverse size of the nucleon for different quark longitudinal-momentum slices.
For details on the GPD formalism, see the
reviews~\cite{Goeke:2001tz,Diehl:2003ny,Ji:2004gf,Belitsky:2005qn,Boffi:2007yc,Guidal:2013rya}.

In the $ep\to e^\prime p^\prime \gamma$ reaction, the DVCS process interferes with the well-known
Bethe-Heitler (BH) process, illustrated in Fig.~\ref{fig:diags} (right), where the final-state
photon is radiated by the incoming or scattered electron.

Extracting the GPDs from the DVCS process requires measuring a series of observables for
the $ep\to e^\prime p^\prime \gamma$ reaction over the broadest
kinematic domain possible. Several
observables, such as the unpolarized cross section and polarized beam or/and target asymmetries, 
are necessary in order to separate the four GPDs. Each 
observable is sensitive to a particular combination of GPDs.

This paper presents a major contribution to this global and long-term endeavour: the extraction of
the $ep\to e^\prime p^\prime \gamma$ (\textit{i.e.}, DVCS+BH) unpolarized and beam-polarized cross
sections over the widest phase space ever explored in the valence-quark region, with 110
($Q^2,x_B,t$) bins covering:
$1.0<Q^2<4.6$~GeV$^2$, $0.10<x_B<0.58$, and $0.09<-t<0.52$~GeV$^2$.
In this kinematic domain, our results strongly enhance the existing set of
measurements of the $ep\to e^\prime p^\prime \gamma$ reaction which consists of:
four ($Q^2,x_B,t$) bins of unpolarized cross sections and 12 bins of beam-polarized cross sections
measured by the JLab Hall A collaboration~\cite{Munoz Camacho:2006hx},
57 bins of beam-spin asymmetries measured by the CLAS collaboration~\cite{Girod:2007aa},
and 18 bins of longitudinal target- and beam-target double-spin asymmetries measured by the
CLAS collaboration~\cite{Seder:2015,Pisano:2015} (in addition to the handful of CLAS pioneering data
points of Refs~\cite{Stepanyan:2001sm,Chen:2006na,Gavalian:2009}).

The experiment took place at JLab during three months in 2005, using the 5.75-GeV polarized
electron beam (79.4\% polarization), a 2.5-cm-long liquid-hydrogen 
target, and the Hall B large-acceptance CLAS spectrometer~\cite{Mecking:2003zu},
operating at a luminosity of 2$\times$10$^{34}$~cm$^{-2}$s$^{-1}$.
A specially designed electromagnetic calorimeter (``inner calorimeter",
IC~\cite{Girod:2007aa}) was added to the CLAS detector and allowed the detection of photons
for polar angles from about 5$^\circ$ to 16$^\circ$, with full azimuthal coverage.
A solenoid magnet was installed around the target in order to magnetically shield the IC from
the copious M{\o}ller background stemming from the target.

The first step of the data analysis was to select events with at least one electron, one proton, and
one photon in the final state.
Electrons were identified by signals in the CLAS drift chambers, scintillators,
Cherenkov counters, and the standard CLAS electromagnetic calorimeters. Protons
were identified by the correlation between their measured momentum and velocity.
The highest-energy particle detected in the IC was considered as a photon candidate.
\begin{figure}[htb]
\vspace{-0.2cm}
\includegraphics[width=9cm]{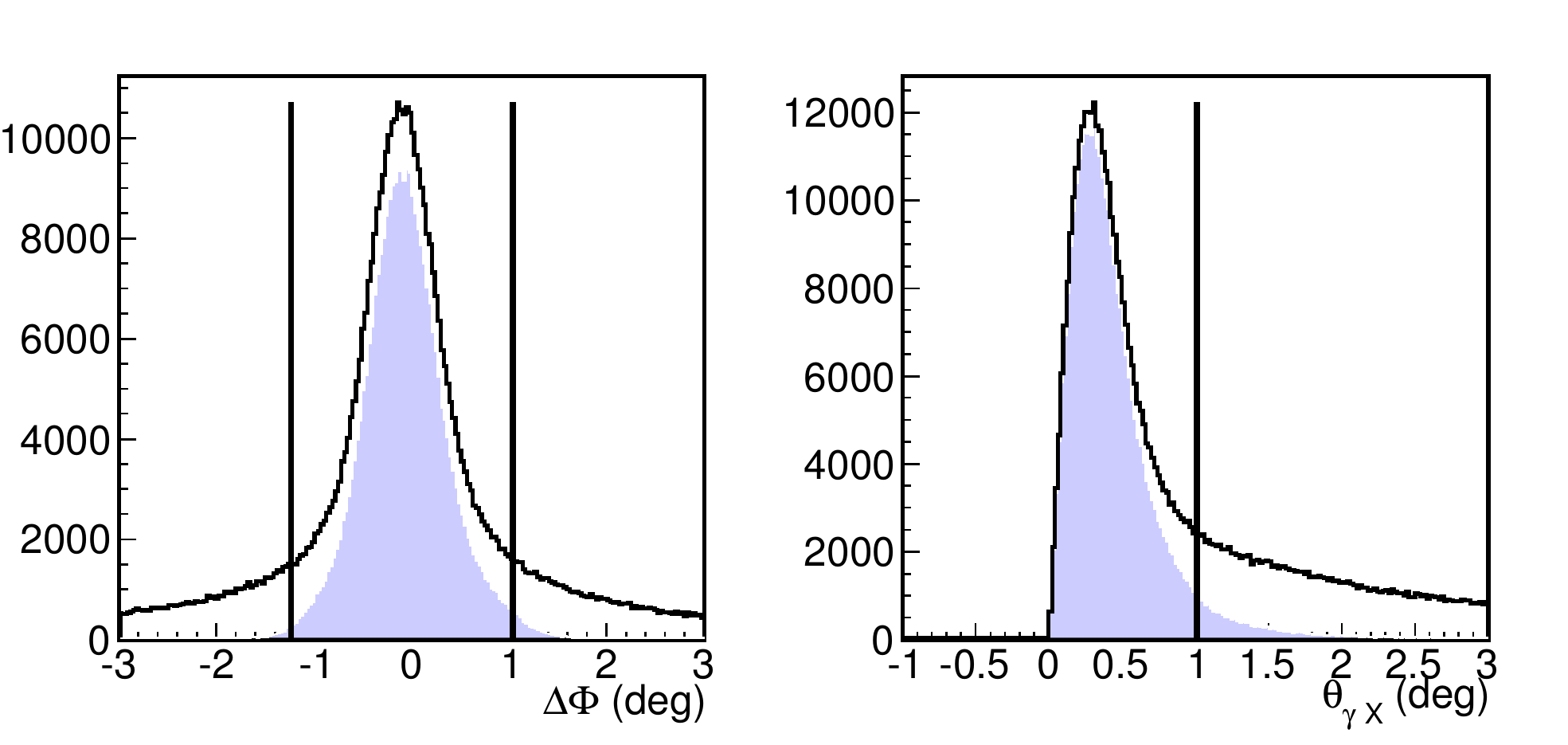}
\vspace{-0.7cm}
\caption{(Color online) Two of the four variables on which $3\sigma$ exclusivity cuts (vertical lines)
were applied to select the $ep\to e^\prime p^\prime \gamma$ reaction: $\Delta \phi$ (left)
and $\theta_{\gamma X}$ (right). 
Black solid distributions correspond to the events with at least one electron and one proton detected
in CLAS, and one photon detected in the IC, after applying the exclusivity cuts on $MM^{2}_{e^\prime p^\prime}$.
Each blue shaded distribution corresponds to the events remaining after
applying the exclusivity cuts on all the variables except for the plotted one.}
\label{fig:cuts}
\end{figure}
Once these three final-state particles were selected and their 3-momenta determined,
the exclusivity of the $ep\to e^\prime p^\prime \gamma$ reaction was ensured by applying $3\sigma$ cuts
on the following four variables: the squared missing mass $MM^{2}_{e^\prime p^\prime}$ of the 
($e^\prime p^\prime X$) system, the coplanarity angle $\Delta \phi$, \textit{i.e.}, the angle between 
the ($\gamma^*,p^\prime$) and ($\gamma^*,\gamma$) planes, the missing transverse momentum of the
($e^\prime p^\prime \gamma$) system, and the angle $\theta_{\gamma X}$ between the measured photon and that
predicted by the kinematics of the ($e^\prime p^\prime X$) system. We also selected the particular kinematics:
$W>2$~GeV, where $W^2=s=(\gamma^*+p)^2$, to minimize contributions from radiative decay of
baryonic resonances, and $Q^2>1$~GeV$^2$ in order to be in the deep virtual regime.
As an example, Figure~\ref{fig:cuts} shows the effect of two of the four exclusivity cuts.

\begin{figure}[htb]
\vspace{-0.3cm}
\includegraphics[width=9cm]{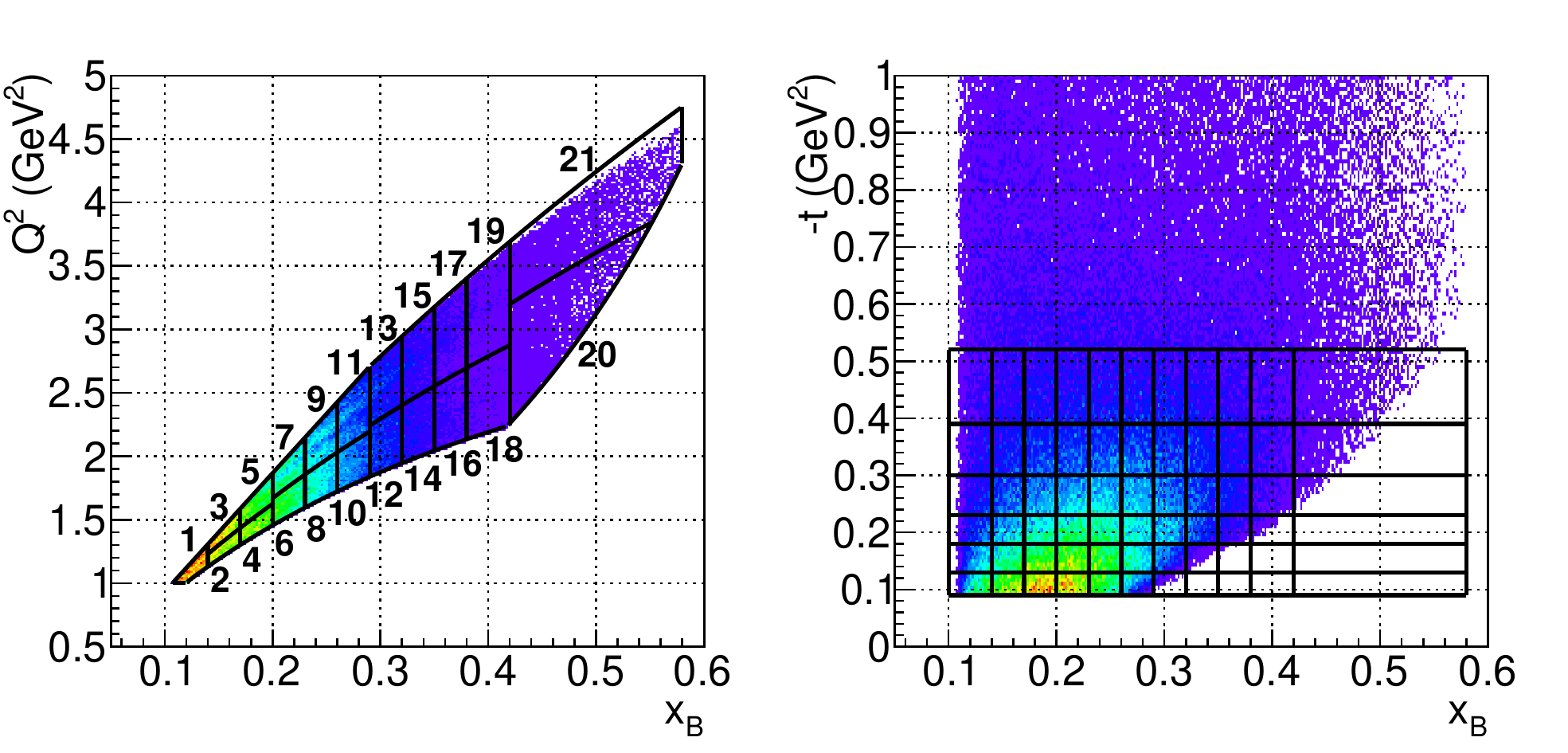}
\vspace{-0.8cm}
\caption{(Color online) The ($Q^2,x_B$) (left) and ($-t,x_B$) (right) kinematic coverages
of this experiment, with the corresponding binning.}
\label{fig:ps}
\end{figure}

Under these conditions, we ended up with about 300,000 events.
Figure~\ref{fig:ps} shows the resulting ($Q^2,x_B$) and ($-t,x_B$) kinematic coverages
of the data and the adopted binning [21 ($Q^2,x_B$) bins and 6 $t$ bins],
which is finer than the one used in Ref.~\cite{Girod:2007aa}.
Note that the bins and results presented here are limited to the $|t|$ region below
0.52~GeV$^2$ while the actual coverage of the data goes beyond 1~GeV$^2$.
The $ep\to e^\prime p^\prime \gamma$ cross sections vary very rapidly
with kinematics,
primarily due to the BH process.
In order to minimize the uncertainties related to the knowledge of the kinematics, 
we chose to minimize the size of our bins, while keeping comparable statistics in
each bin.

Due to the azimuthal symmetry when using an unpolarized target,
the $ep\to e^\prime p^\prime \gamma$ reaction depends on four independent variables.
For the study of GPDs, the most appropriate ones are $Q^2$, $x_B$, $t$ and
$\phi$, where $\phi$ is the azimuthal angle between the ($e,e^\prime$) 
and ($\gamma^*,p^\prime$) planes around the virtual photon direction.
We have thus extracted four-fold cross sections as follows:
\begin{equation}
\frac{d^4 \sigma_{ep \rightarrow e^\prime p^\prime \gamma}}{dQ^2 dx_B dt d\phi} =
\frac{N_{ep\to e^\prime p^\prime \gamma}}{\mathcal{L}_{int} \Delta Q^2 \Delta x_B \Delta t \Delta \phi
\ Acc \ F_{rad}}.
\label{eq:sig}
\end{equation}
In Eq.~\ref{eq:sig}, 
$N_{ep\to e^\prime p^\prime \gamma}$ is the number of $ep\to e^\prime p^\prime \gamma$ events in the
($Q^2,x_B,t,\phi$) bin. The aforementioned exclusivity cuts do not fully select a pure
sample of DVCS+BH events. We evaluated the contamination from the $ep\to e^\prime p^\prime \pi^0$
channel where one photon of the $\pi^0$ decay can escape detection, using a combination of
$ep\to e^\prime p^\prime \pi^0$ measurements and Monte-Carlo simulations.
On average, this contamination is less than 9\% and was subtracted on a bin-by-bin basis.
The four-dimensional acceptance/efficiency of the CLAS detector, $Acc$, for the
$ep\to e^\prime p^\prime \gamma$ reaction
was determined for each ($Q^2,x_B,t,\phi$) bin by generating more than 200 million
DVCS+BH events, using a realistic Monte-Carlo generator. The events were processed through 
the GEANT simulation of the CLAS detector, and the same reconstruction
and analysis codes that were used for the data.
The event generator includes radiative effects so that $Acc$
also corrects for a part of the real internal radiative effects.
The factor $F_{rad}$ corrects, for each ($Q^2,x_B,t,\phi$) bin, for the virtual internal
radiative effects and the remainder of the real internal radiative effects, which can be both
calculated theoretically~\cite{Akushevich:2012tw}.
The product ($\Delta Q^2 \Delta x_B \Delta t \Delta \phi$) corresponds to the effective
hypervolume of each bin.
Finally, $\mathcal{L}_{int}$ is the effective integrated luminosity, corrected for the data
acquisition dead time, which was deduced from the integrated charge of the beam measured by a
Faraday cup. In addition, we applied a global renormalization factor of 12.3\%, determined
from the analysis of the elastic scattering $ep\to e^\prime p^\prime$, by comparing the
experimental cross section to the well-known theoretical one. This factor compensates for
various kinematic-independent inefficiencies, not well reproduced by the simulations.

\begin{figure}[htb]
\includegraphics[width=8.7cm]{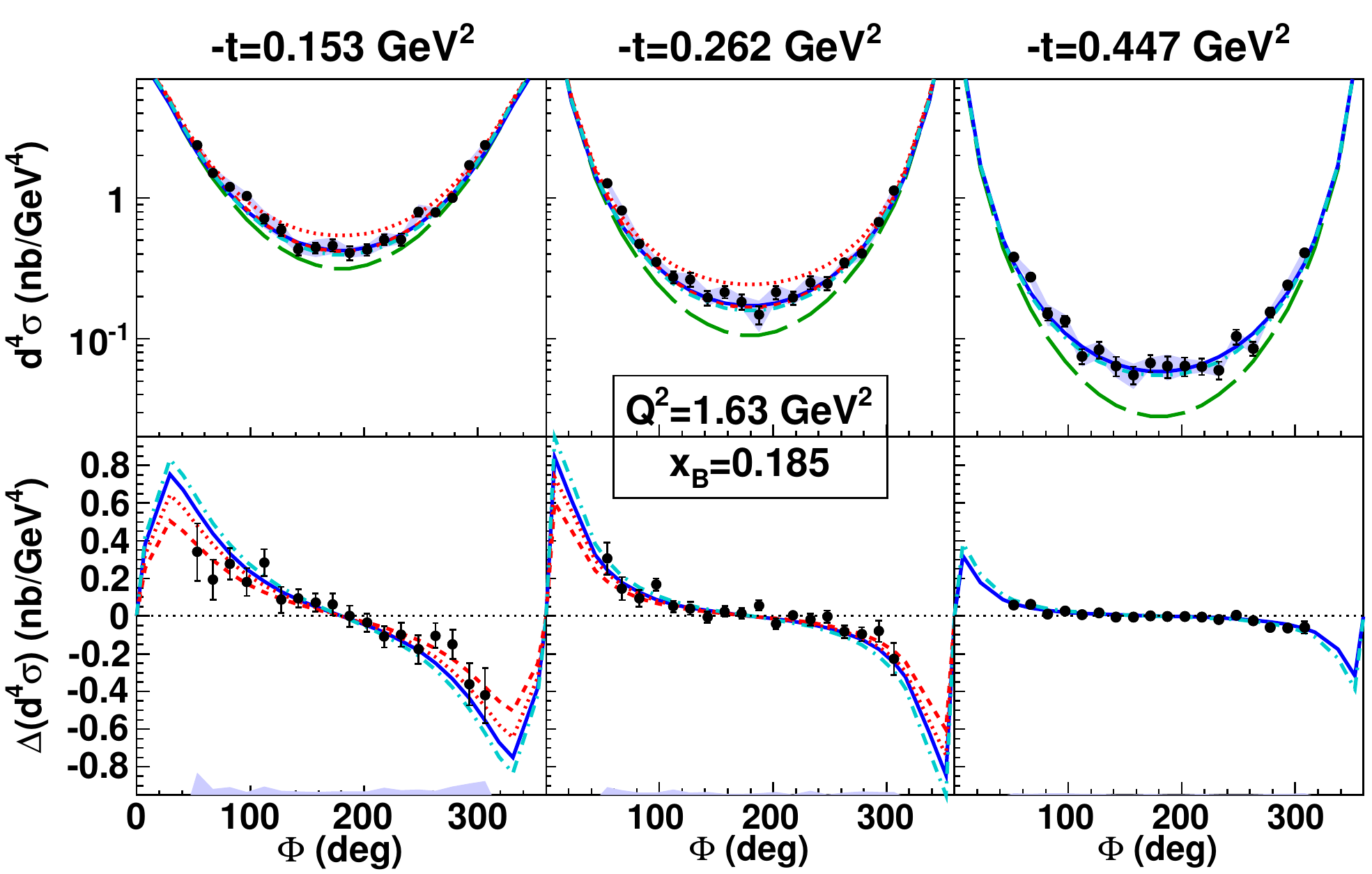}
\includegraphics[width=8.7cm]{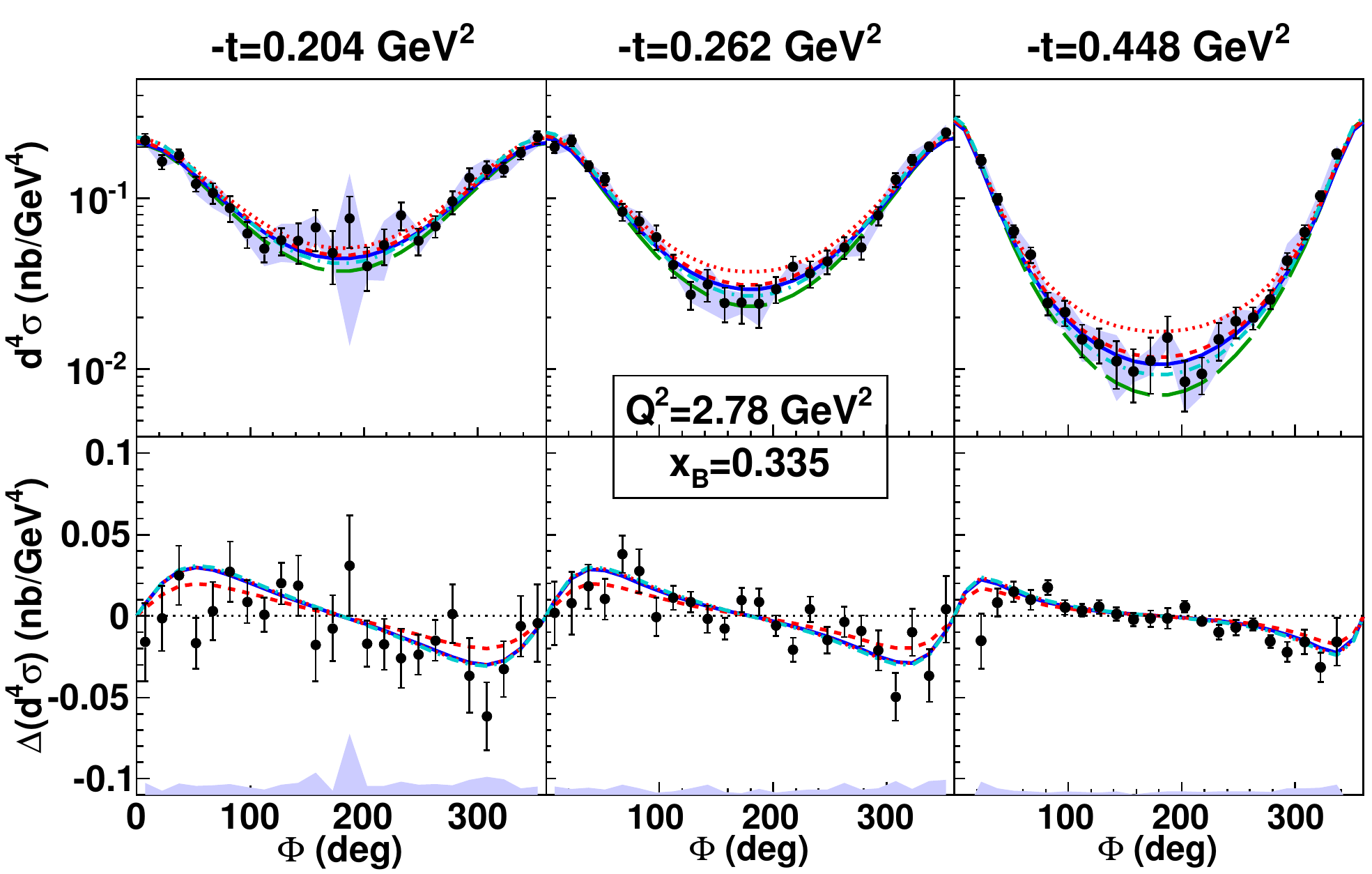}
\vspace{-0.5cm}
\caption{(Color online) Top six plots: unpolarized cross section 
$\frac{d^4 \sigma_{ep\to e^\prime p^\prime \gamma}}{dQ^2 dx_B dt d\phi}$ (top row)
and beam-polarized cross-section difference $\Delta(d^4\sigma)$ for the
$ep\to e^\prime p^\prime \gamma$ reaction, as a function of $\phi$,
for ($Q^2,x_B$)=(1.63 GeV$^2$, 0.185) and for 3 $-t$ values: 0.153, 0.262 and 0.447~GeV$^2$.
Bottom six plots: same observables for ($Q^2,x_B$)=(2.78 GeV$^2$, 0.335) and 
$-t$=0.204, 0.262 and 0.448~GeV$^2$.
The green long-dashed curves show the BH contribution only.
The other curves correspond to the predictions of four GPD models:
VGG~\cite{Vanderhaeghen:1998uc1999xj,Goeke:2001tz,Guidal:2004nd} 
(blue solid curves), KMS~\cite{KMS} (cyan dash-dotted curves),
and two versions of the KM model~\cite{Kumericki:2009uq,Kumericki:2011zc},
KM10 (red dotted curves) and KM10a (red short-dashed curves). The blue bands show the
systematic uncertainties.}
\label{fig:xsec}
\end{figure}

Figure~\ref{fig:xsec} shows, for two selected ($Q^2,x_B$) bins in different parts
of the phase space, the $\phi$-dependence of the $ep\to e^\prime p^\prime \gamma$
unpolarized cross section and beam-polarized cross-section difference.
The latter of these two observables is defined as follows:
\begin{equation}
\Delta(d^4\sigma)=\frac{1}{2}\left[\frac{d^4 \overrightarrow{\sigma}_{ep\to e^\prime p^\prime \gamma}}
{dQ^2 dx_B dt d\phi}
-\frac{d^4 \overleftarrow{\sigma}_{ep\to e^\prime p^\prime \gamma}}{dQ^2 dx_B dt d\phi}\right],
\end{equation}
where the arrows correspond to beam helicity states $+$ and $-$.
For each of these ($Q^2,x_B$) bins, three selected $t$ bins are shown.
In Fig.~\ref{fig:xsec},
the black error bars show the statistical uncertainties of the data [13.9\% on the unpolarized
cross section on average, over the 110 ($Q^2,x_B,t$) bins] and the blue bands show the
systematic uncertainties [14\% on the unpolarized cross section on average].
The contributions to the latter include the uncertainties on the beam energy and
therefore the kinematics and associated corrections (5.7\% on average), 
the acceptance correction (5.3\%), the global renormalization factor (5\%),
the exclusivity cuts (3.5\%), the radiative corrections (2.2\%), the particle selection
(1.6\%), and the $\pi^0$ background subtraction (1\%).

The unpolarized cross sections peak towards
$\phi$=0$^\circ$ and 360$^\circ$ due to the BH process for which the final-state  
photon is predominantly emitted in the direction of the initial or scattered electron.
This is quantitatively confirmed by the calculations shown in Fig.~\ref{fig:xsec},
where the green curves show the BH contribution only. The difference between the BH
curves and the data can thus be attributed to the DVCS process, and therefore linked to GPDs. 
We display in Fig.~\ref{fig:xsec} calculations of four GPD models, listed in the caption.
The modeling of the GPDs in the VGG and KMS models is based on the Double-Distribution 
representation~\cite{Radyushkin:1998es,Radyushkin:1998bz,Mueller:1998fv}. The VGG
calculations in Fig.~\ref{fig:xsec} only include the contribution of the GPD $H$
as the inclusion of the other GPDs barely changes the results. The KM model is based 
on the Mellin-Barnes 
representation~\cite{Kumericki:2009uq, Mueller:2005ed}. The KM10 version of the 
model includes contributions from all four GPDs for which the free parameters
were fitted to the JLab~\cite{Munoz Camacho:2006hx,Girod:2007aa}, 
HERMES~\cite{Airapetian}
and ZEUS/H1~\cite{Chekanov:2003ya,Aktas:2005ty} data.
In that work, it was found that it is possible to fit the JLab Hall A unpolarized cross sections
only at the price of the introduction of a very strong $\tilde H$ contribution~\cite{Htilde}.
The KM10a version is based on a fit which excludes the JLab Hall A unpolarized cross sections~\cite{Munoz Camacho:2006hx} 
and sets $\tilde H$ to zero. Note that none of these four models has been tuned to our data.

Figure~\ref{fig:xsec} shows that the predictions of standard GPD models like VGG, KMS, and KM10a,
whose compatibility is remarkable in spite of their different approaches,
are in good agreement with our unpolarized cross-section data. In contrast, we see that the
KM10 version, which includes the strong $\tilde H$ contribution, tends to overestimate our data.
Over our 110 ($Q^2,x_B,t$) bins, the average $\chi^2$ value
per degree of freedom~\cite{chi2}
is 1.91 for VGG, 1.85 for KMS, 
1.46 for KM10a, and 3.94 for KM10.
We can therefore conclude that standard GPD models with a dominant contribution of the
GPD $H$ to the unpolarized cross section, \textit{i.e.}, without the introduction of a strong
$\tilde H$ contribution, describe the data well. Moreover, the disagreement between our data
and the KM10 model, which instead matches the Hall A results, might reveal an inconsistency
between
the two sets of data. As a check, we performed a dedicated data analysis using the exact same
($Q^2,x_B,t$) bin limits as those used for the Hall A analysis ($Q^2$=2.3~GeV$^2$, $x_B$=0.36,
and $-t=$0.17, 0.23, 0.28 and 0.33~GeV$^2$). However, in this limited and particular
($Q^2,x_B,t$) region, the comparison is hampered by our large statistical
uncertainties and lack of $\phi$-coverage around $\phi=180^\circ$.
Thus no conclusion can be drawn from this comparison.
The Hall A experiment was run at a luminosity almost
three orders of magnitude larger than ours, but in a much more limited phase space.

In general, the four models, including KM10, give a good description of the beam-polarized
cross-section difference and the data barely allow one to distinguish one model from another.
Over our 110 ($Q^2,x_B,t$) bins, the average $\chi^2$ value
per degree of freedom~\cite{chi2}
is 1.40 for VGG, 1.84 for KMS,
1.06 for KM10a, and 1.20 for KM10. 

Finally, we attempted to extract directly some GPD information from these two sets of observables.
We used the local-fitting procedure developed in
Refs.~\cite{Guidal:2008ie,Guidal:2009aa,Guidal:2010ig,Guidal:2010de}.
At leading-twist and leading-order, this procedure uses well-established DVCS amplitudes and
does not depend on model parametrizations of the GPDs. We fit simultaneously the
$\phi$-distributions of our unpolarized and beam-polarized cross sections
at a given ($Q^2,x_B,t$) kinematic point by the eight (real) quantities:
\begin{eqnarray}
F_{Re}(\xi,t) &=& {\cal P}\int_{-1}^{1}dx\left[\frac{1}{x-\xi}\mp\frac{1}{x+\xi}\right]
F(x,\xi,t), \nonumber
\end{eqnarray}
\begin{eqnarray}
F_{Im}(\xi,t) &=& F(\xi,\xi,t) \mp F(-\xi,\xi,t). \label{def_cffs}
\end{eqnarray}
In Eq.~\ref{def_cffs}, $F=H, \tilde H, E,\tilde E$, the top and bottom signs apply to the
unpolarized ($H,E$) and polarized ($\tilde H,\tilde E$) GPDs respectively, and ${\cal P}$ is the
principal value integral.
These quantities are called Compton Form Factors (CFFs)~\cite{CFFdef2} in
Refs.~\cite{Guidal:2008ie,Guidal:2009aa,Guidal:2010ig,Guidal:2010de}
and ``sub-CFFs" in Ref.~\cite{Kumericki:2013lia}.
The only model-dependent input in the procedure is that the CFFs are
allowed to vary in a very conservative limited range,
$\pm$5 times the CFFs from the VGG model~\cite{Guidal:2004nd}.
In spite of the underconstrained nature of the problem, \textit{i.e.}, fitting two
observables with eight free parameters, the algorithm manages in general to find well-defined
minimizing values
for $H_{Im}$ and $H_{Re}$. The reason is that the two observables
that we fit are dominated by the contribution of the GPD $H$.

\begin{figure}[tb]
\includegraphics[width=8.7cm]{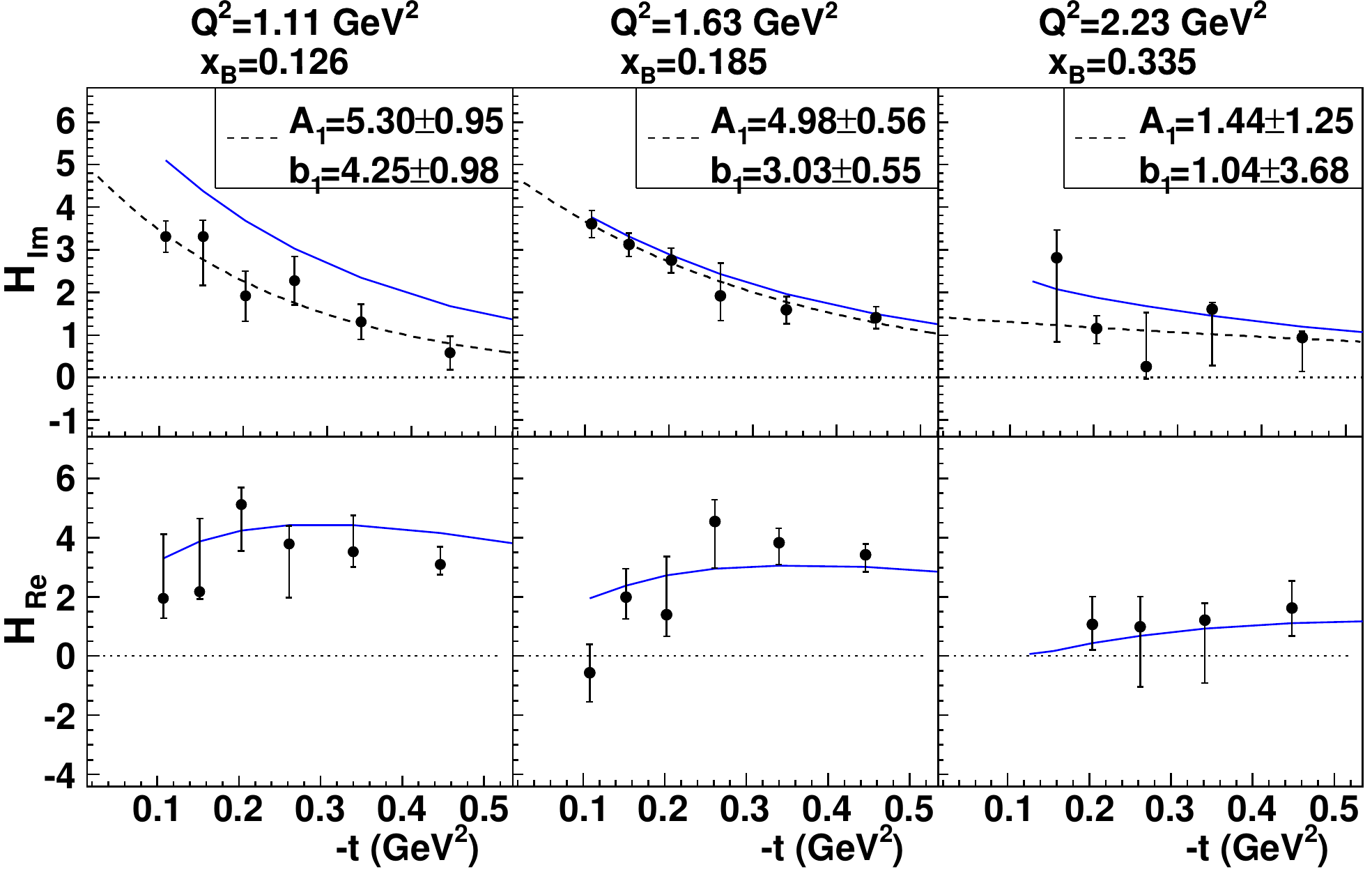}
\vspace{-0.7cm}
\caption{(Color online) Results of the CFF fit of our data for
$H_{Im}$ (upper panel) and $H_{Re}$ (lower panel), with only
the GPDs $H$ and $\tilde H$, for three of our ($Q^2,x_B$) bins, as a function of $t$.
The blue solid curves are the predictions of the VGG model.
The black dashed curves show the fit of the results by the function $Ae^{bt}$.}
\label{fig:cff}
\end{figure}

Ideally, one would like to fit all CFFs. However, with only two observables in this case,
this leads to too large uncertainties. We therefore present in Figure~\ref{fig:cff}, for
a selection of three of our 21 ($Q^2,x_B$) bins, the $t$-distribution of the fitted
$H_{Im}$ and $H_{Re}$, computed neglecting the contributions associated with $E$ and
$\tilde E$.
Fig.~\ref{fig:cff} also shows the predictions of the VGG model, which
overestimates the fitted $H_{Im}$ at the smallest values of $x_B$.

We have fitted, in Fig.~\ref{fig:cff}, the $t$-dependence of $H_{Im}$ by
the function $Ae^{bt}$ with the normalization $A$ and the slope $b$ as free parameters.
Keeping in mind that the $Q^2$ values are different
for the three $x_B$ bins, the results of these fits show 
that $A$ and $b$ increase, in a systematic way, with decreasing $x_B$.
Under the hypothesis of neglecting $Q^2$ higher-twist and evolution effects  
as well as \textit{deskewing} effects~\cite{deskewing},
these behaviors might reveal tomographic features of the quark content of the nucleon. 
Under the mentioned conditions, $b$ is related to the transverse size of the nucleon.
Our data therefore suggest, over the $x_B$ range explored in this experiment, that the size of
the nucleon increases as lower momentum fractions (proportional to $x_B$) are probed.
The rising of $A$ reflects the increase of the partonic content of the nucleon
as lower $x_B$ values are probed.
$H_{Re}$ does not lend itself easily to a simple interpretation
as it involves a weighted-integration of the GPD $H$ over $x$. Nevertheless, its
extraction is of great use to constrain models.

In conclusion, we have measured the unpolarized and beam-polarized four-fold cross sections
$\frac{d^4 \sigma}{dQ^2 dx_B dt d\phi}$
for the $ep\to e^\prime p^\prime \gamma$ reaction, for 110 ($Q^2,x_B,t$) bins, over the
widest phase space ever covered in the valence-quark region. This large set of new data
will provide stringent constraints on GPD models. We have shown that three
well-known GPD models, VGG, KMS, and the KM10a version of the KM model, describe
the data well without additional inputs.
The model interpretation of the present results favors a smaller deviation from the pure
BH process around $\phi=180^{\circ}$ than suggested by the Hall A data.
Within such models, this reinforces the expectation of the $H$-dominance in the
unpolarized cross section.
We have also extracted the $H_{Im}$ and $H_{Re}$ CFFs from our data, fitting simultaneously
both cross-section observables. Under some assumptions, our results tend to show that the
nucleon size increases at lower parton-momentum values, thus revealing a
first tomographic image of the nucleon.

The full data set is available at~\cite{CLASdata}.
A long article is in preparation and will include the results for all our
($Q^2,x_B,t$) bins and a more detailed description of the data analysis.

We thank the staff of the Accelerator and Physics Divisions
at Jefferson Lab for making this experiment possible.
We also thank I.~Akushevich, K.~Kumeri\v{c}ki and D.~Mueller
for informative discussions and making available their calculations.
This work was supported in part by
the U.S. National Science Foundation,
the Chilean Comisi\'on Nacional de Investigaci\'on Cient\'ifica y Tecnol\'ogica (CONICYT),
the French Centre National de la Recherche Scientifique (CNRS),
the French Commissariat \`{a} l'Energie Atomique (CEA),
the French-American Cultural Exchange (FACE),
the Italian Istituto Nazionale di Fisica Nucleare (INFN),
the National Research Foundation of Korea (NRF),
the Scottish Universities Physics Alliance (SUPA),
and the United Kingdom's Science and Technology Facilities Council (STFC).
This work benefited from the support of the French Agence Nationale de la Recherche
(contract ANR-12-MONU-0008-01 PARTONS) and the Joint Research Activity GPDex of
the European program Hadron Physics 3 under the Seventh Framework Programme of the European
Community. This material is based upon work supported by the U.S. Department of Energy,
Office of Science, Office of Nuclear Physics under contract DE-AC05-06OR23177.


\begin{thebibliography}{99}
\bibitem{Htilde}
The introduced $\tilde H$ contribution is about a factor 3 larger than values given by
standard parametrizations, such as in VGG, or inferred from simple relations to
polarized Parton Distribution Functions.
\bibitem{CFFdef2}
The CFFs are alternatively defined as four complex quantities whose real and imaginary
parts are respectively as: 
$\Re{\rm e}{\cal F}(\xi,t) = F_{Re}(\xi,t)$ and 
$\Im{\rm m}{\cal F}(\xi,t) = -\pi F_{Im}(\xi,t)$.
\end{thebibliography}
\end{document}